\documentclass{ws-mpla}
\usepackage[usenames,dvipsnames]{xcolor}
\usepackage[numbers]{natbib}
\usepackage[caption=false]{subfig}
\usepackage{graphicx}

\begin{document}

\markboth{Mathews, et al.}
{Origin and Evolution of Structure and Nucleosynthesis for Galaxies in the Local Group}

\catchline{}{}{}{}{}

\title{ORIGIN AND EVOLUTION OF STRUCTURE AND NUCLEOSYNTHESIS FOR GALAXIES IN THE LOCAL GROUP}

\author{\footnotesize G. J.  MATHEWS, A. SNEDDEN, L. A. PHILLIPS, I.-S. SUH, J. COUGHLIN, A. BHATTACHARYA, }

\address{Center for Astrophysics, University of Notre Dame, Department of Physics\\
Notre Dame, IN 46556,
USA
gmathews@nd.edu
asnedden@nd.edu
Lara.A.Phillips.127@nd.edu
isuh@nd.edu
jcoughl2@nd.edu
abhatta2@nd.edu}

\author{X.ZHAO}

\address{Department of Astronomy \& Astrophysics, Penn State University, University Park, PA 16802, USA
xuz21@psu.edu}

\author{N. Q. LAN}

\address{Hanoi National University of Education, 136 Xuan Thuy, Hanoi, Vietnam
nquynhlan@hnue.edu.vn}

\maketitle


\begin{abstract}
The Milky Way  is the product  of a complex evolution of generations of mergers, collapse, star formation, supernova and collisional heating, radiative and collisional cooling, and ejected nucleosynthesis.  Moreover, all of this occurs in the context of the cosmic expansion, the formation of cosmic filaments, dark-matter halos, spiral density waves, and emerging dark energy.  In this review we summarize observational evidence and discuss recent  calculations concerning  the formation, evolution, and nucleosynthesis in the galaxies of the Local-Group.  In particular, we will briefly summarize  observations and simulations for the dwarf galaxies and the  two large spirals  of the Local Group.  We discuss how  galactic halos form within the  dark matter filaments that define a super-galactic plane.  Gravitational interaction along this structure leads to streaming flows  toward the two dominant  galaxies in the cluster.    These simulations and observations also suggest that  a significant fraction of the Galactic halo formed as at large distances and then arrived later along these streaming flows. We also consider the insight provided by observations and simulations of nucleosynthesis both within the galactic halo and dwarf galaxies in the Local Group.

\end{abstract}

\ccode{PACS Nos.: 98.35.-a, 98.35.Ac, 98.35.Bd, 98.35.Hj, 98.52.Wz, 98.56.-p, 98.56.Si, 98.56.Wm}

It has been clear for some time [e.g.~\cite{White78}] that the Milky-Way Galaxy did not form  in isolation as the collapse of a single cloud as was originally proposed \cite{Eggen62}.  Rather it is the product of the development of a much more extended structure and the chaotic merging of smaller structures \cite{Searle78}.  This structure begins within the initial dark-matter potentials formed during the radiation-dominated epoch and then evolves into the filament/void morphology characteristic of the standard cosmological constant plus cold-dark-matter ($\Lambda$CDM)  cosmology.  Within this structure there has been a complex sequence of heating by mergers,  star formation and supernovae, along with collisional and radiative cooling  and the collapse of star forming cold molecular clouds. One must analyze all of these processes within the entire extended early Local Group  (LG) in order to understand the properties of the Milky Way and its satellite systems. 

   Indeed, it is these satellite systems that give the best glimpse into the early evolution of the LG, as these systems are remnants of the initial merging protogalactic structures that were to form the present-day dominant large galaxies of the LG.  It is the goal of this review to summarize some of the current understanding and challenges regarding the formation and evolution of the galaxies and nucleosynthesis in the LG.  A great deal of evidence has been compiled recently toward this understanding while at the same time, large scale simulations are approaching sufficient resolution to present a plausible history of how the local galactic environment  has come into being.  
   
   Exactly where the LG begins and ends is ambiguous as our system extends all the way to the Virgo cluster in what makes up the local super group. A reasonably updated of members and positions has been maintained by the SEDS organization \cite{LGmembers}.  The LG is dominated by two large galaxies, plus  45 definite members including dwarf spheroidals, irregular galaxies, dwarf ellipticals, a couple of SBm Magellanic spirals.  A list of 37 possible members has also been added to that group of 45 members.  It is the dwarf  galaxies in particular that we wish to discuss  in this review, though as we will see, it is impossible to separate their origin from that of the larger members of the LG. Properties of dwarf spheroidals include \cite{Strigari08} an absolute magnitude of $M_V \sim -17$  compared to    $M_V \sim -20.5$ for the Milky way.  Moreover, as discussed below in this review, it is now known [e.g. \cite{Gilmore07}] that these galaxies can be  dominated by dark matter. In $\Lambda$CDM cosmology, these dwarf members are, in part, the building blocks of the Galaxy.

   One puzzling, but revealing aspect of the LG is the fact that galaxies in the LG exhibit an alignment.  This  has been known since the pioneering work of  Kunkel \& Demers \cite{Kunkel76}  and Lynden-Bell \cite{ Lynden-Bell76,Lynden-Bell82}, where it was noted that the galaxies of the LG are aligned in a great circle. Moreover, subsequently discovered  satellite galaxies in the Sloan Digital Sky Survey ({\it SDSS}) [Ref.~\cite{Koposov08} and references therein]  appear also to be aligned upon the great circle of the Local Group \cite{Kroupa05,Metz09,Kroupa10}. The same phenomenon has been also found for M31 \citep{Grebel99,Hartwick00,Metz09}.  Furthermore, some of the satellites and even globular clusters have been shown to have a coherent motion \cite{Lynden-Bell95,Metz08,Keller12,Pawlowski12}. It has been noted \cite{Godlowski10} that on scales of $\le 10$ Mpc there is a tendency for galaxies to align in a direction toward the local supercluster.
   
   A number of works  [e.g.~\cite{Libeskind05,Libeskind10, Zentner05, Kang05,Deason11,Lovell11,Wang12a}] have performed  numerical simulations in an attempt to explain this in the current $\Lambda$CDM model. It has been found, however,  that the possibility to have such an alignment, and especially one with coherent motion in the LG is low. It has been also argued \citep{Zwicky56,Lynden-Bell76,Metz08} that the alignment of the satellites in the LG may come from a tidal origin. This scenario has also been studied with numerical simulations  \cite{Fouquet12,Pawlowski11,Pawlowski12} illustrating  that one can easily reproduce a similar alignment to the one found in the LG. However, if this is the case, the mass-to-light ratio of these tidal dwarf galaxies in the simulations cannot match the high values that have been observed \citep{Simon07} since they tend to only have gas in them \citep{Bournaud10}.
   
   In our own work \cite{Mathews12}, we have extended these investigations to analyze numerically the generic tendency for LG-type poor clusters to exhibit an alignment and streaming motion toward the dominant galaxies of the cluster.  Our simulations  imply that the LG should contain 1 to 3 aligned systems currently inflowing.  We discuss this below.   
   
\section{Numerical Simulations}

It is  straightforward [e.g. \cite{Mathews12,Zhao11a,Zhao11b}] to do  large scale structure simulations with random initial conditions  in a standard $\Lambda$CDM cosmology.  One begins by specifying the content, e.g..  $\Omega_\Lambda = 0.726$ and  $\Omega_M = 0.274$ and a baryon content of $\Omega_B = 0.0456$ as deduced from the Wilkinson Microwave Anisotropy Probe {\it WMAP} \cite{WMAP} or Planck Surveyor \cite{Planck}.  One problem \cite{Gomez12} with trying to reconstruct a model for the LG is that, the determination of "best-fit" parameters is not unique. Indeed, very different halo merger histories can reproduce the same observational data set, . Thus, attempts to uniquely characterize the formation history of the LG using numerical simulations must be done statistically  by  analyzing large samples of high-resolution N-body simulations.

The code that we and others have adopted for such numerical simulations of large scale structure and LG formation  is the n-body smoothed-particle hydrodynamics (SPH) code GADGET, originally developed by Springel et al. \cite{Springel01}. The current public version GADGET-2  \cite{Springel05a} is used for all of the simulations specifically shown in this review, although a newer  GADGET-3 Version now exists and is used by some researchers in the field.
  
Perhaps, the most ambitious recent applications of this code have been the Aquarius project \cite{Springel08}. The Aquarius Project  consists of  cosmological
simulations of the formation of six dark matter haloes with
a mass and merging history similar to that expected for the halo of the Milky Way. For the highest resolution level in the simulations, the main halo contains about 1.5 billion particles with nearly 3$\times 10^5$ sub-halos in it. The high mass resolution run utilizes a particle mass of  $10^ 3 M_{\odot}$making it possible to study the ultra faint satellite galaxies in the simulations.
The simulations assume a $\Lambda$CDM cosmology, with a matter density parameter of $\Omega_M = 0.25$;
and cosmological constant of $\Omega_\Lambda = 0.75$.  A matter power spectrum normalization of 
$\sigma_8  = 0.9$ with spectral slope  of $n_s = 1$, and a Hubble parameter of $h = 0.73$  were adopted.  

Two large collaborations were established to study the effects of adopting different modeling techniques \cite{Kim14} and feedback prescriptions \cite{Scannapieco12} on simulations of a Milky Way sized galaxy in a Local Universe-like environment. The AGORA Project \cite{kim14} uses common astrophysics packages and initial conditions in order to study the effects of adopting alternative codes such as the smoothed particle hydrodynamics code GASOLINE \cite{Wadsley04,Stinson06}, or adaptive mesh refinement codes such as ART \cite{Kravtsov97}, ENZO \cite{Bryan95, Enzo13}, and RAMSES \cite{Teyssier02} on the resulting evolution of dwarf galaxies in cosmological simulations and in particular on a model of a Milky Way sized galaxy  and satellites in a Local Universe-like environment. The Aquila Comparison Project \cite{Scannapieco12} contrasts thirteen gas-dynamical simulations of a Milky Way galaxy embedded in a Local Universe and, in particular, finds that the effect of the adopted feedback on the prominence of the stellar disk overwhelms the variations due to the modeling technique. The statistics of satellite galaxies around Milky-Way like hosts have been studied in the Millennium simulation \cite{Springel05b,Wang12} where the brightest satellites were determined to be more massive than those found in observations.   Also the Bolshoi simulation \cite{Busha11} who finds find agreement with SDSS observations in the probably of a Milky-Way-like halo having none, one, or two Magellanic Cloud-like satellites.   A summary of links to various projects and codes is given in the Appendix.

The good news is that all of these different simulation codes essentially
agree in terms of dark matter dynamics and non-radiative gas physics as long as the resolution is sufficient.  On the other hand, the bulk properties of simulated galaxies
depend almost entirely on the details of star formation
and feedback of the heating from stars and supernovae.  Prescriptions for this vary widely from code to code [e.g. \cite{Stinson06}].  Moreover,  most simulations either lack altogether or have problems with: black hole feedback; detailed chemical evolution of various elements; effects of dust; magnetic fields; radiation transport (ionization balance); and the detailed mixing of ejected elements.  Ultimately all simulations invoke some level of approximation and only by detailed comparison with observations can the validity of the approximations be tested.   

In our own studies \cite{Mathews12,Zhao11a,Zhao11b} we have modified the public version of the GADGET-2 code \cite{Springel05a} to incorporate  our own cooling, star formation and supernova feedback routines.  We have even considered  relativistic corrections to the conformal Newtonian gauge \cite{Zhao11a,Zhao11b}, but found such corrections to be both cumbersome and negligible. To construct a cooling/heating rate table dependent upon redshift, density, metallicity and temperature, we used the code {\it Cloudy v10.0},  last described by Ferland et al.~\cite{Ferland98}.  This code accounts for  collisional and  Compton ionization,  UV-photoionization, two photon, continuum,  metal cooling and free-free scattering.  Since the cooling timescales are much smaller than the simulation time steps, we also implicitly integrate the interpolated radiative heating / cooling contribution to the entropy equation (cf.~\cite{Springel02}).

We have developed our own method to account for the total energy, mass, and metals ejected from supernovae Types I and II.  We start with the method described in the chemo-dynamical model of  \cite{Kobayashi04} to compute the feedback from supernovae and nucleosynthesis.   The supernovae metal enrichment fraction of the ejected mass was interpolated from the tables presented by Nomoto et al. \cite{Nomoto97a,Nomoto97b}.  The subsequent energy and mass ejected was then distributed to nearby gas particles, in a weighted manner, using the standard SPH smoothing kernel.  To prevent the problem of the distributed energy being radiating away immediately \cite{Katz92}, we adopted the method of \cite{Stinson06}.  All gas particles within the Sedov blast wave have their radiative cooling turned off for 30My.  This gives the distributed energy enough time to thermalize and allow the gas particles to adiabatically expand.

 
In simulations without star formation or supernovae feedback, the deduced temperature vs. density relation for gas particles in our simulation of a Local-Group-like system  is similar to those from other simulations  [e.g.~ \cite{Katz96, Springel03}].  That is, we reproduce the expected  broad temperature distribution for gas densities comparable to the background density and a population of gas up to high density with temperatures near $10^4$ K due to the Lyman $\alpha$ edge.



We have made numerous simulations with different random initial conditions on volumes of ($15-400$ Mpc h$^{-1}$) $^3$.   For each large scale simulation, an initial linear matter power spectrum  was generated using the online interface\footnote{$http://lambda.gsfc.nasa.gov/toolbox/tb_camb_form.cfm$} of the CAMB code which is based on the standard CMB algorithm CMBFAST \cite{Lewis00, Seljak96}.  This power spectrum was then  used to generate random non-Gaussian fields using second-order Lagrangian perturbation theory.  For this purpose we  utilized the  2LPTic\footnote{$http://cosmo.nyu.edu/roman/2LPT/$} \cite{Scoccimarro12,Crocce06}.  The subsequent initial-condition was  generated in a Gadget-2 ready format. 

To find suitable initial conditions with a Local-Group-like poor clusters, we ran many low resolution simulations (2 $\times$ 32$^3$ particles) on a $15$Mpc/h$^3$ box and used a friends-of-friends algorithm \cite{Liu08} to identify two close Milky-Way  \& Andromeda-sized galaxies in the simulations.  Then the simulations were run again at higher resolution (2 $\times$ 128)$^3$ and (2 $\times$ 256)$^3$ particles to identify the dwarf galaxies unresolved in the low-resolution runs.  These higher resolution runs have modest effective gas particle mass resolutions of $2.8\times10^7M_{\odot}$ and $3.6\times10^6M_{\odot}$, respectively.  Thus, it might be necessary to perform follow-up runs in a smaller box with better particle resolution once an LG-like system is successfully identified.

\begin{figure}[htb]
\centering
\includegraphics[angle=-0, width=15cm]{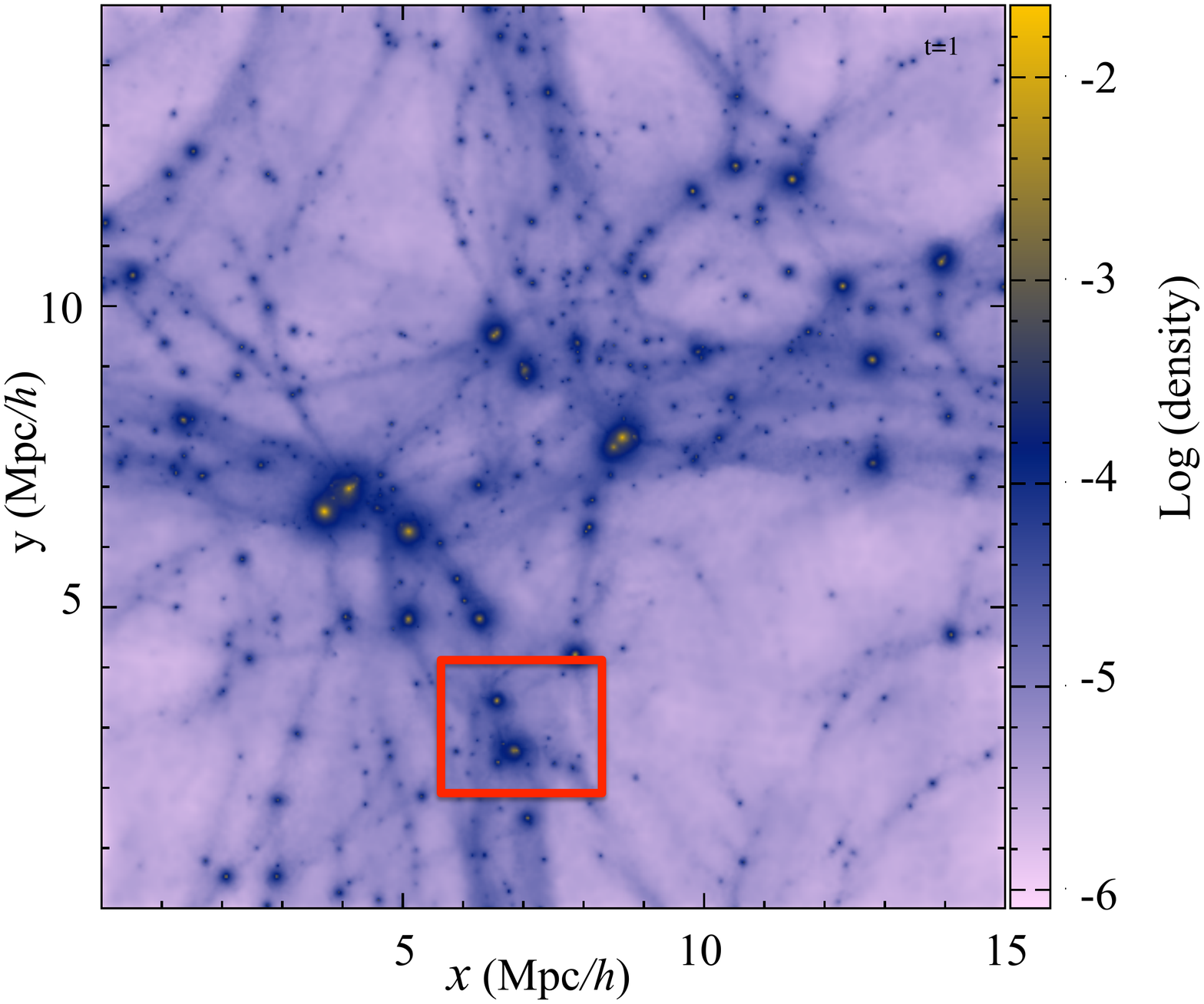}		
\caption{Log of baryonic column-density distribution in the $x-y$ plane for a large scale 15 Mpc cosmological simulation.  The box in the lower portion of the figure identifies  a LG-like system  characterized by  two large galaxies separated by $\sim 800$ kpc and surrounding halo galaxies.\label{LSS-15}}
\end{figure}

Figure \ref{LSS-15} shows an example of a 15 Mpc large scale simulation in which a LG-like system can be seen as two small galaxies at the bottom of the simulation box separated from a local supercluster at a distance (somewhat closer than that of the LG) of about 5 Mpc/h.  

\begin{figure}[htb]
\centering
\includegraphics[angle=-0, width=15cm]{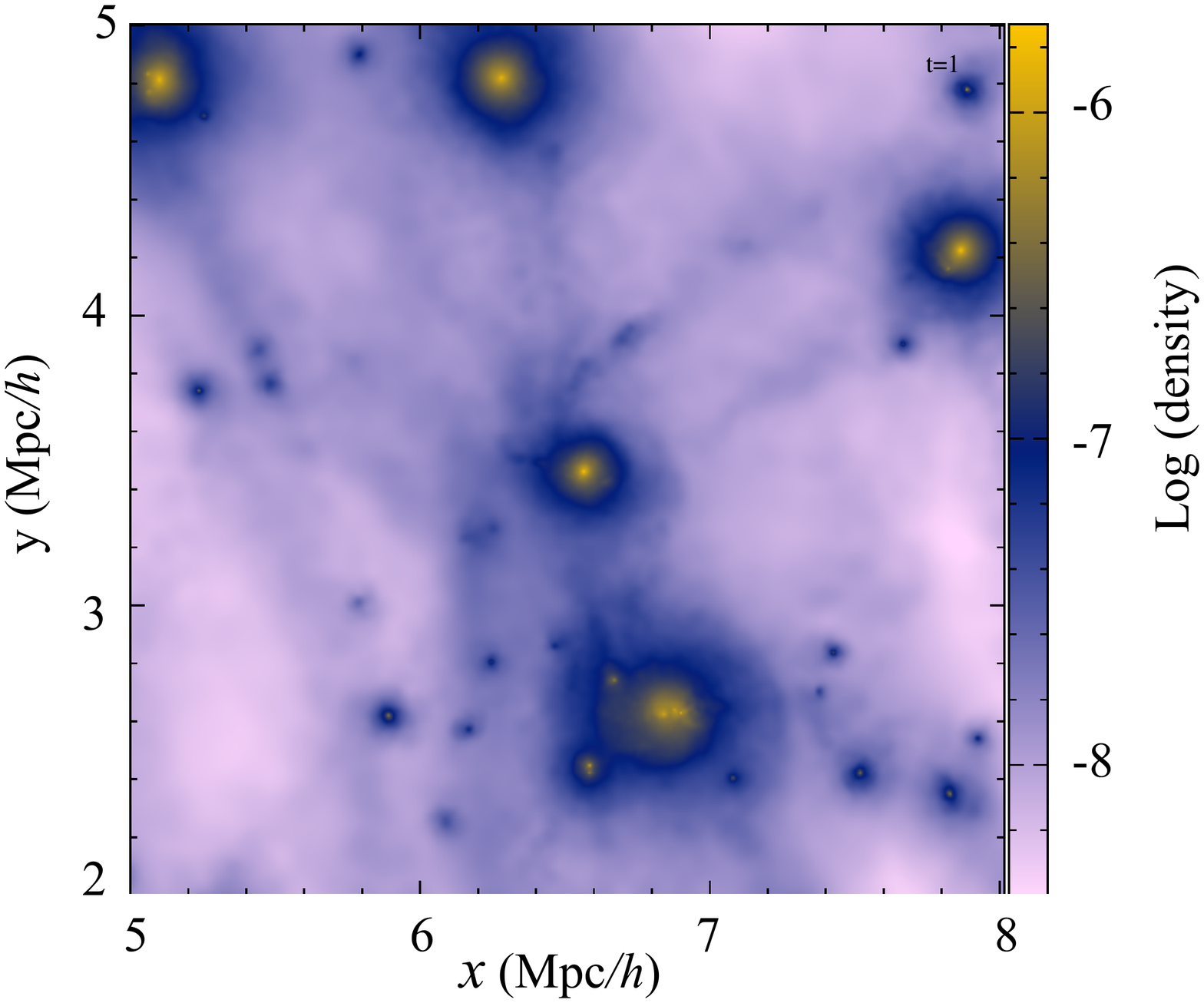}		
\caption{Log of column density distribution in the $x-y$ plane for a on a scale of 3 Mpc for the LG-like system in the  cosmological simulation of fig.~\ref{LSS-15}. \label{LG-3}}
\end{figure}

Figure \ref{LG-3}  shows a zoom to a scale of 3 Mpc/h for this LG system identified in Fig. \ref{LSS-15}. Note the numerous dwarf galaxies surrounding the two dominant galaxies.  Note also, that the largest dwarf galaxies tend to reside in filamentary structures that generate a streaming flow toward the largest galaxies \cite{Mathews12}.

\section{Simulations of Local-Group Like Systems}
 Of particular interest \cite{Mathews12}  is the formation of a large number of infalling dark matter halos and/or dwarf galaxies neatly aligned along the filamentary structure of the simulation. Indeed, it is clear from the simulations that  the stellar halo of the dominant galaxies is, at first,  the result of a number of infalling dwarf galaxies forming and merging very early, but subsequently a continuous stream of dwarf galaxies arriving up to much later times and  from distances of up to hundreds of kpc away.

 When the baryons are included, the two dominant spiral galaxies of masses $\sim1-4 \times 10^{11}$ M$_\odot$ form early in the simulation.  Also the presence of Magellanic-cloud or M32-like systems and  {\it SagA}- like mergers occurring near the Galactic center is a common feature of such systems and easily identifiable on Figure \ref{LG-3}.   
 
\begin{figure}[htb]
\centering
\includegraphics[angle=-0, width=12cm]{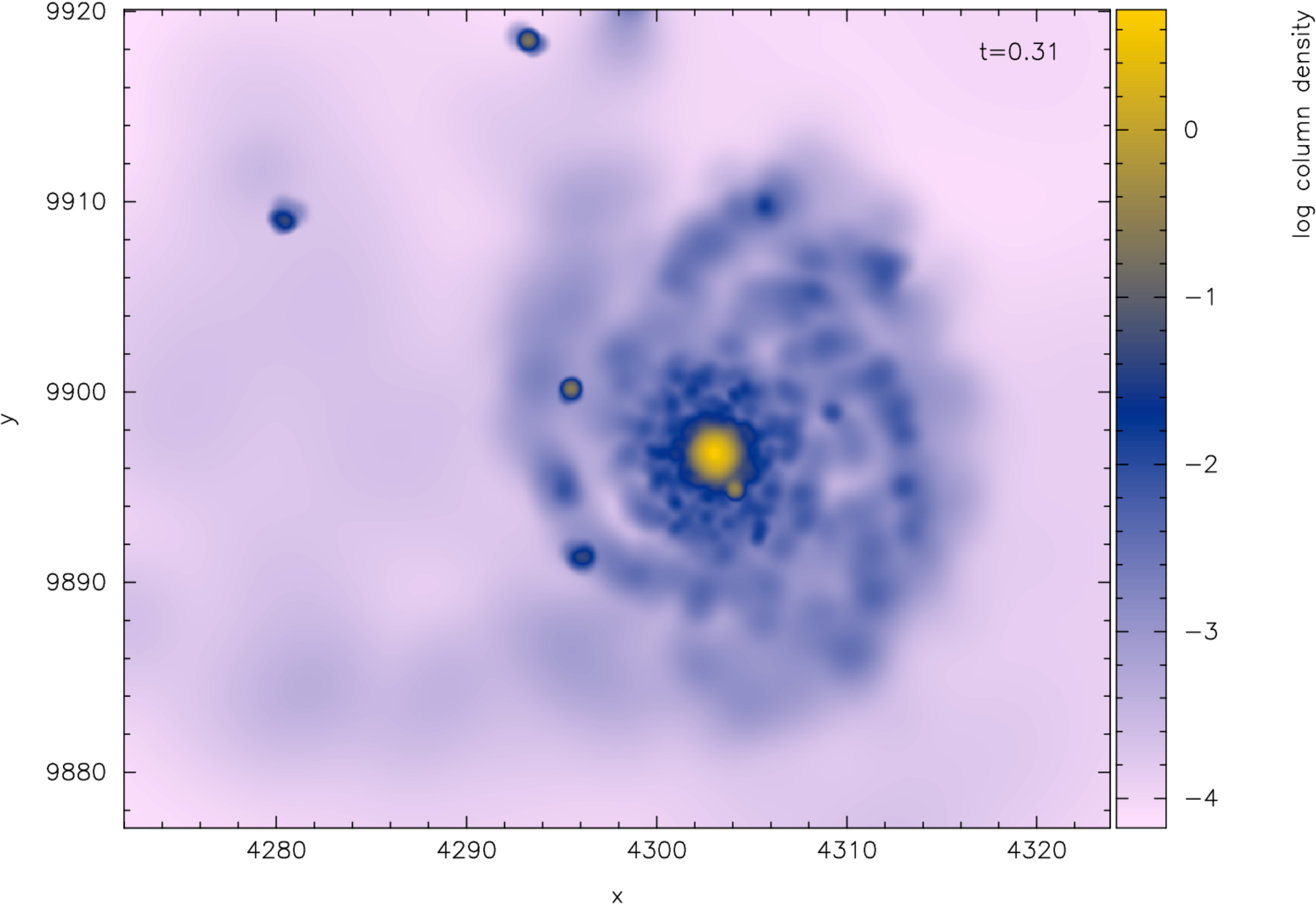}		
\caption{Density distribution in the $x-y$ plane for A Milky-Way-like spiral galaxy formed at redshift z=2.2 in one of our simulations. The length scale is in units of $\rm {h^{-1} kpc}$. The density is in units of $ \rm {10^{10} h^{-1} M_{\odot}/ (h^{-1} kpc)^3}$.\label{galaxy_pdf}}
\end{figure}

\begin{figure}[htb]
\centering
\includegraphics[angle=-0, width=6cm]{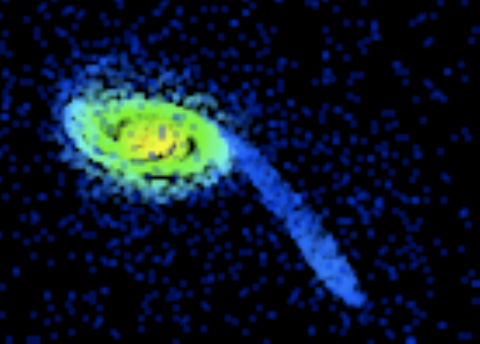}
\includegraphics[angle=-0, width=6cm]{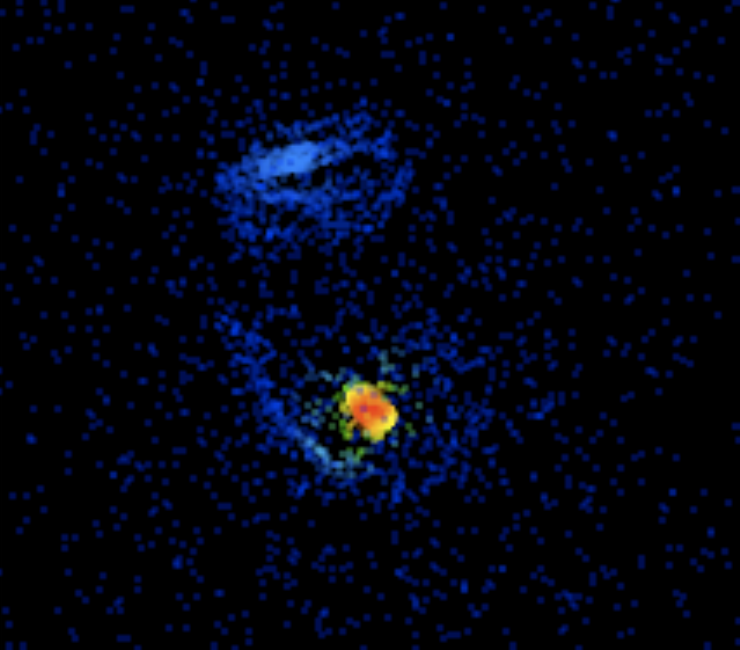}
\includegraphics[angle=-0, width=6cm]{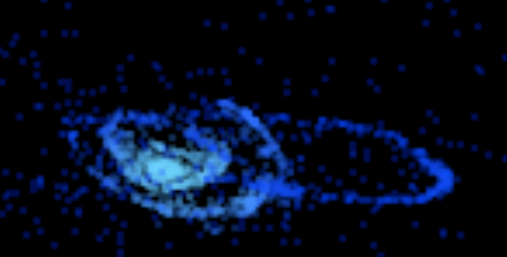}	
\caption{Images (at z=0) of dwarf spiral galaxies from the simulation.  Note the streaming flows and tidal distortion due to recent mergers.\label{magellanic_pdf}}
\end{figure}

 Figure \ref{galaxy_pdf} shows an example of a Milky-Way like spiral from one of our simulations \cite{Zhao11b}, while Figure \ref{magellanic_pdf} shows examples of magellanic-like SBm galaxies from another  simulation.   Of particular interest in these simulations is that the inflowing dwarf  galaxies   line up along  filamentary  structures.  This alignment  causes an enhanced gravity along the filament leading to a streaming flow of cosmic fluid  toward the dominant galaxies.   
 
\begin{figure}[!htb]
\centering
\includegraphics[width=5in,clip]{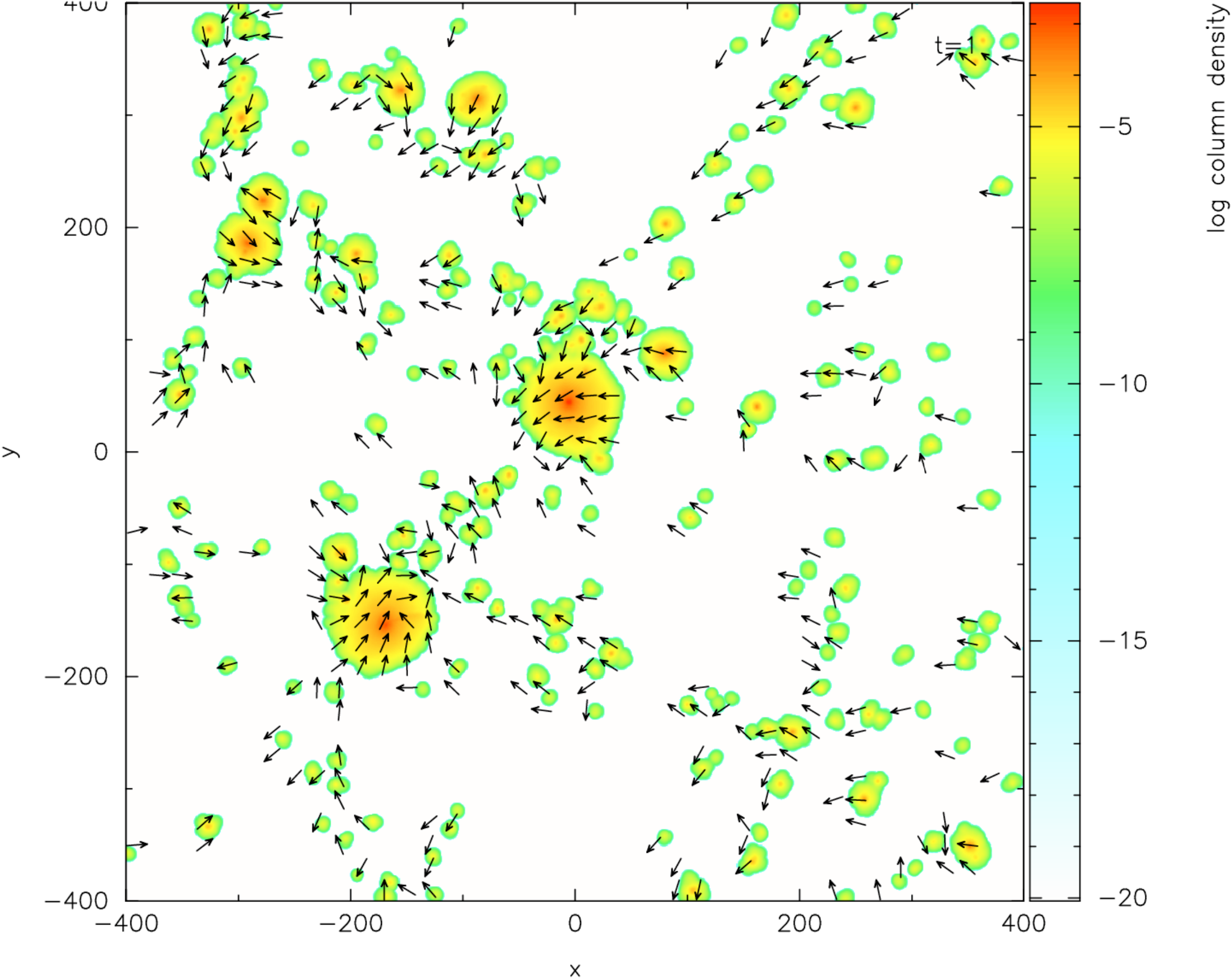}
\caption{Distribution of  streaming flows in the XY plane toward two dwarf galaxies in one of the simulations. Arrows drawn indicate the direction of comoving flow.  
Note the appearance of  three streaming flows;  one from the lower right, one from the upper left, and one from the upper right. \label{position-2D-sim}}
\end{figure}

 Figure \ref{position-2D-sim} shows an example \cite{Mathews12} of a pair of Magellanic type galaxies from one of our LG simulations.  Note the appearance of at least three  streaming flows;  one from the lower right, one from the bottom,  and one from the upper right.  In larger systems these  correspond to a flow of arriving/merging dwarf spheroidal galaxies  from hundreds of kpc throughout the simulation history.  The implication for the LG system is that a significant fraction of the most distant ($> 1$ Mpc) observed dwarf galaxies should line up into such flows.  Indeed,  there there is  marginal evidence \cite{Mathews12} in observed dwarf galaxies  for such streaming flows arriving from in the general direction of the Moffie and Sculptor systems (though the Sculptor galaxy itself is probably not a part of this motion).  However, more data on dwarf galaxies at large distances is desired.

From the simulations one can  deduce that many of the "first stars" could have formed far from the Milky Way and later arrived via streaming flows. Indeed, late time mergers are the likely source of an observed population \cite{Beers12} of counter orbiting low-metallicity stars in the outer halo. 

\section{Observations of Metal-Poor Stars in Dwarf Galaxies}

 From the $\Lambda$CDM simulations described above one expects that the dwarf spheroidal galaxies of the LG arrive along streaming flows.  Indeed, there is observational evidence not only of streaming flows in the Milky-way \cite{Lehner11, Barger12,Fox13}, but there is also evidence \cite{Martinez12} of streaming flow toward the dwarf galaxy NGC4449.
 
Observed properties of dwarf spheroidals include \cite{Strigari08} an absolute magnitude of $M_V \sim -17$ compared to $M_V \sim -20.5$ for the Milky way.  Moreover, it is now known \cite{Gilmore07} that these galaxies are dominated by dark matter. In $\Lambda$CDM cosmology, these galaxies are the building blocks of the Galaxy. As such, stellar metallicities and abundances for metal-poor stars in dwarf galaxies should agree with those in the  halo of the Milky Way.
Extremely metal poor (EMP) stars in dwarf galaxies could provide detailed  information on the formation and early chemical evolution of the Galaxy.
As discussed below, a persistent problem, however, has been that the number of observed dwarf galaxies is less than expected from numerical simulations. As discussed in Sec. 5, this missing satellite problem [e.g. \cite{Moore99b}] may result if many of the expected dwarf galaxies are of low surface brightness and not yet detected or may be an artifact of the simulations.  Also, as we now summarize, there is a lack of the expected very low metallicity stars in the halo dwarf galaxies.

\subsection{Chemical Abundance Trends}
Abundance studies for individual stars in dwarf galaxies is a topic of intense current interest [e.g. \cite{Tolstoy09,Aoki12,Frebel11,Frebel12}].  Although there are some similarities, in general the chemical abundance trend of (metal-rich) dwarf spheroidal galaxies is different from that of the Milky way halo.  In particular there is a trend \cite{Tolstoy09} toward diminished abundances for $\alpha$ elements (O, Mg, Si, Ca) relative to the Milky way  for $-2 <$ [Fe/H] $< 0$.
This could suggest a large contribution  from SNIa or a tendency to eject material emitted from SNII's out of the shallow gravitational potential of dwarf galaxies relative to deeper gravitational potential of the Milky way.  

A good clue to to which of these is the correct interpretation may be found in the search for neutron capture elements in dwarf galaxies.  In \cite{Letarte10} abundances of neutron-capture  elements Ba, Y, La, Nd and Eu were determined in Fornax.   Significant excesses of s-process  [Ba/Fe], [La/Fe], and [Nd/Fe] were found for many stars were found relative to those of the Milky way.  In particular Ba, the element best representing the second s-process peak for [Fe/H]$\sim -1$ shows a consistent enhancement of [Ba/Fe] relative to the Milky Way by as much as 1 dex.   While the nearly pure r-process element Eu is nearly identical to the Milky Way.  This is significant since it may indicate that the slower s-process ejecta in AGB winds remains trapped in dwarf galaxies relative to the more energetic supernova production of Fe and r-process elements.

Regarding the s-process, yttrium is an s-process element   near the first s-process peak.  However, the [Y/Ba] ratio of this element to another s-process element, barium,  is diminished \cite{Tolstoy09} in Fornax relative to the Milky Way.  Since Y is mainly the result of a weak s-process in massive stars while Ba is produced in the second s-process by the main s-process in AGB stars, this again supports the case that the supernova ejecta from massive stars is launched out of  the ISM of  dwarf spheroidals.

Alpha elements in dwarf spheroidals have also provided useful insight \cite{Aoki12,Frebel12,Hill07}.  In Ref.~\cite{Hill07} it is noted that [Ca/Fe] is suppressed relative to the Milky Way not only for Fornax, but also in the Sgr and LMC dwarf galaxies.   This again supports the idea dwarf galaxies do not retain supernova ejecta.

Another general observation \cite{Helmi06,Winnick03,Martin07, Koch08a, Koch08b, Kirby08} is that there is an overall lack of stars with metallicity [Fe/H] $<-3$ in dwarf galaxies.   This suggests the  possibility that dSphs are not really the left over  building blocks of early galaxies, since one should find the remnants of the first low metallicity stars within them.  A possible solution to this dilemma, however is suggested by the discovery \cite{Okamoto08} of  a population of faint dwarf galaxies with very low metallicity in the Sloan Digital Sky Survey (SDSS) archive.  
In addition to lower metallicity compared with brighter dwarf galaxies, these ultra-faint dwarf galaxies have absolute magnitudes similar to typical globular clusters, but have much larger size.   This population not only solves the missing first stars problem but has a good chance to solve the missing satellite problem as well \cite{Wang12b} (see below).

\section{Remaining Challenges in the Simulations}
Although simulations of Local-Group like systems have been attempted for decades, a few persistent challenges have remained.
There are three problems that are so prevalent in the simulations that they have acquired names.  These are the core-cusp problem, The missing-satellites problem, and the too-big-to-fail problem.  There is also the problem of the delicate balance between stellar/SN feedback and cooling/star formation, and the problem of poor resolution for dwarf galaxies, and the lack of a galactic bulge in the LG.  These problems we now describe in detail. 

\subsection{Core-Cusp Problem}
The core-cusp  problem remains as one of the unresolved challenges  between observation
and simulations in  the standard  $\Lambda$CDM model for the formation of the LG and galaxies in general.
Basically, $\Lambda$CDM simulations predict that the center of galactic dark matter halos contain  a steep power-law mass density profile  \cite{Navarro97, Moore99a, Ishiyama11}. However, observations of dwarf galaxies in the LG  \cite{Moore94,Burkhert95,Oh11}
consistently reveal a  density profile that is a nearly flat  distribution of  dark matter near the 
the center.  

A number of solutions to this dilemma have been proposed.  One possibility
 \cite{Mashchenko08} is  that periodic variations in the galactic potential driven by
episodic star burst events may flatten the central dark-matter density profile. In particular, these bursts of star formation
would induce a largeÐscale outflow from the galactic center and flatten the density profiles.  This mass loss leads to gas
depletion and temporarily terminates the star formation. As the ejected gas later falls back
toward the galactic center, a large amount of internal energy is lost by radiative cooling.  
After  sufficient cold gas has re-accumulated, another  starburst can occur. This repeated star-burst induced expansion and
contraction of the interstellar gas could lead to a cyclic change in the gravitational potential
around the center of galaxies and flatten the density profile of dark-matter halos. 

Indeed, some
 simulations \cite{Mashchenko08,Governato10,Pontzen12} have exhibited this episodic starburstÐoutflow cycle along with a
core-cusp transition at the center of the halos. In \cite{Ogiya11}, however  it was argued  that mass loss driven by stellar feedback may not 
be an effective mechanism to flatten
the central cusp.  Nevertheless, in \cite{Ogiya13} the importance of  the recurrence frequency of star formation  in determining the dynamical response of 
dark-matter halos was studied.  A comparison was made between the  numerical simulations and  an analytical model of the resonance between 
dark-matter  particles and
the density waves of the interstellar gas. This analysis supports the hypothesis  that there is a resonance between dark-matter particles and the gas that 
 is most effective in flattening the CDM density profiles and resolving the coreÐcusp problem.

\subsection{Missing Satellites Problem}
\begin{figure}[htb]
\centering
\includegraphics[angle=-0, width=10cm]{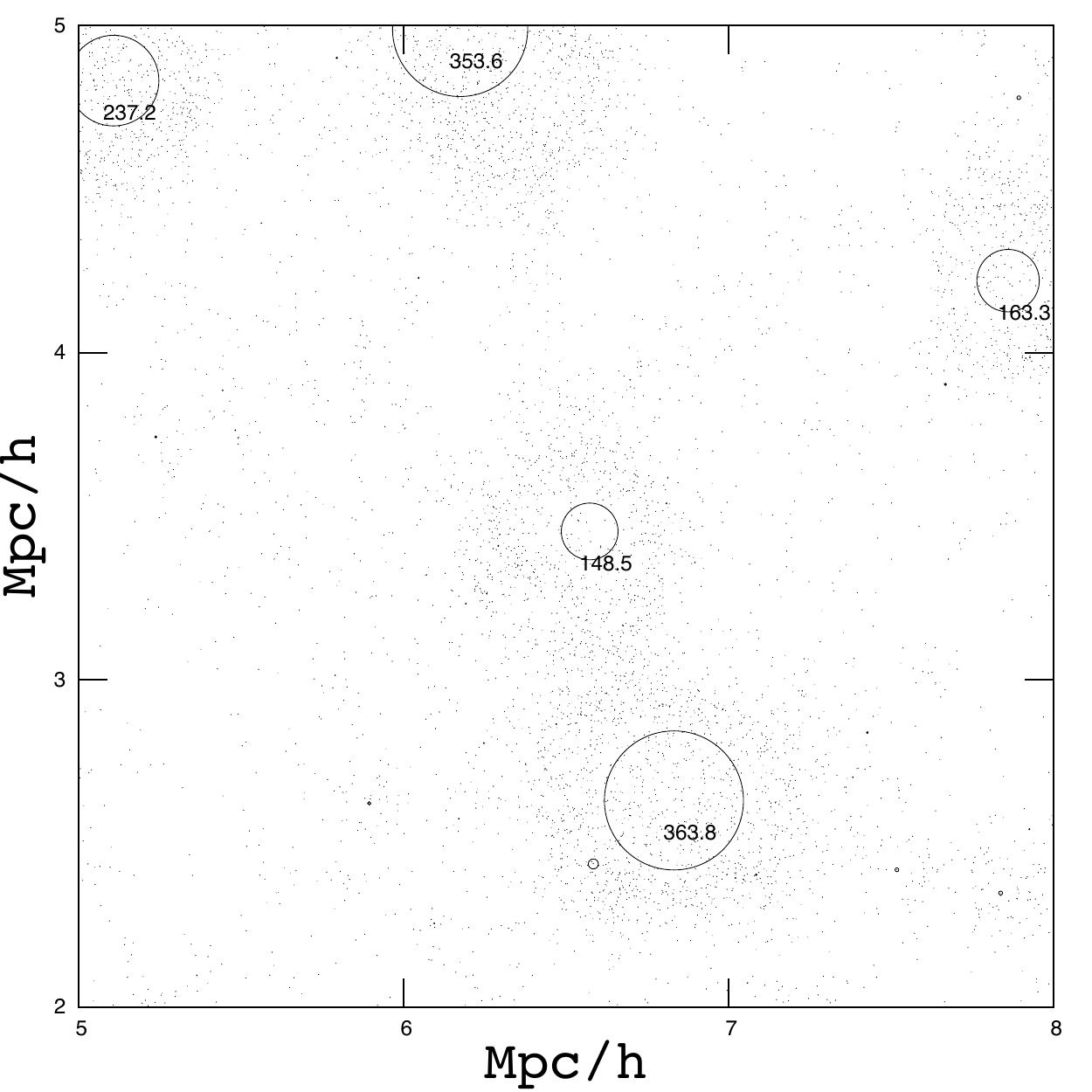}		
\caption{Identification of halos in the XY plane from the LG simulation shown in Figure \ref{LG-3}.  Circles are drawn around identified halo galaxies with a diameter proportional to the enclosed mass.  [Note, the circles do not indicate the region enclosed by the galaxy.  Numbers in the larger circles indicate the total (baryon plus dark-matter) mass in units of $10^{10}$ M$_\odot$
	The two larges circles indicate a Milky Way and Andromeda like spirals.  The smallest points on the plot correspond to systems with $\sim 10^5$ M$_\odot$.    This simulation identifies $> 500$ dSphs within a 3 Mpc box.\label{halos_pdf}}
\end{figure}

The missing satellites problem, also known as the dwarf galaxy problem is the observation that the number of observed dwarf galaxies \cite{mateo98} is orders of magnitude below expected   from dynamical $\Lambda$CDM simulations [e.g.~\cite{Moore99b,Wang12}].  Numerical $\Lambda$CDM simulations generically  predict the hierarchical clustering  and an ever increasing number counts for smaller and smaller sized proto-galactic halos. Although the distribution of normal-sized galaxies in most simulations seems reasonable, the observed number of dwarf galaxies \cite{mateo98} is orders of magnitude lower than expected from simulations \cite{Moore99b}.  For example there are  only about 50  dwarf galaxies observed in the LG \cite{LGmembers}, and only  11 of those orbit around the Milky Way \cite{mateo98}.  On the other hand,   dark matter simulations predict \cite{Moore99b} $\sim  500$ dwarf satellites orbiting the Milky Way alone.  Figure \ref{halos_pdf} from one of our simulations illustrates this.  dots and circles drawn on this figure correspond to galactic halos with M$\ge 10^5$ M$_\odot$.  There are clearly many more than 500 such halos predicted in the simulation.

There are a number of possible solutions to this problem. One is that the smaller halos actually exist but most of them have low baryon content, and hence, low luminosity.  In support of this hypothesis, Keck observations \cite{Simon07} have discovered eight  ultra-faint Milky Way dwarf satellites and showed that six were comprised of  99.9\% dark matter (with a mass to light ratio of about 1000) \cite{Simon07}.   Another solution may be that dwarf galaxies tend to be tidally disrupted by the larger galaxies. This tidal stripping would make it difficult to identify dwarf galaxies, since these objects have low surface brightness and are highly diffuse.

 Another intriguing possible solution \cite{Saitoh13} is that there may be a fundamental flaw in the shock-capturing formulation of the SPH  simulations themselves.
 The standard SPH formulation assumes that the local
density distribution is differentiable. This assumption is used to derive the spatial derivatives of
physical variables. However, this assumption breaks down at a shock contact discontinuity.  It can be demonstrated \cite{Saitoh13} that for standard
 SPH  the density on the low-density side can be  overestimated while that on the high-density side
is  underestimated. Because of this, the pressure of the low (high) density side is over (under) estimated.
This ultimately leads to an unphysical repulsive force at the contact discontinuity, resulting in an effective surface
tension. It is this surface tension that suppresses fluid instabilities and can lead to excessive clustering in merger simulations. 
 
 In \cite{Saitoh13} a new formulation of
SPH has been proposed  in which the internal energy density (pressure) instead of density,  are smoothed quantities
at the contact discontinuity.  This new formulation has been dubbed "density independent SPH (DISPH)." 
 Its employment may alleviate some of the excessive clustering seen in existing SPH simulations,
 although this does not resolve the problem also seen, for example,  in Eulerian AMR codes.

\subsection{Too big to fail problem}
Independent of the dark matter
density profile of the  dSphs in the LG, it is now clear
that the dark matter masses of the dSphs are significantly
lower than that expected from  $\Lambda$CDM simulations.  This is called the "too-big-to-fail" (TBTF)
 problem \cite{Boylan-Kolchin11, Boylan-Kolchin12a}. The
core-cusp and too-big-to-fail problems may be  related. 
If most of the bright dSphs in the LG
have dark matter cores that extend from 0.5  to 1 kpc, then the flattening of their central
cores would also have reduced their masses by a factor of 2-3 compared to a cuspy core.  That is  the
amount required to explain the TBTF problem. 

If, however,  a large fraction of dSphs have non-flattened  profiles, however, the two issues
become  distinct. 
Although this appears as a plausible solution to the
core-cusp problem, it has been pointed out \cite{Charbonnier11} that the brighter dwarf galaxies and low-surface-brightness galaxies  
 can have mass-to-light ratios in excess of 100.  Hence, they are very dark matter dominated. Moreover, at these mass scales one expects the stellar mass to drop
by $\sim$2.5 dex for a difference of only one decade in halo
mass \cite{Behroozi12}. One therefore expects that the dark matter gravitational potential
should overwhelm that of the baryons, even at the center of halos. Finally, the fact that these systems are so deficient
in stars implies that there is not much energy available in supernovae and stellar ejecta to alter the
gravitational potential.

It is noteworthy, however, that some simulations  [e.g. \cite{Governato12, DelPopolo12}] have successfully obtained low central densities
for low-surface-brightness galaxies by means of supernova feedback.   It has also been demonstrated [e.g. \cite{Read05, Zolotov12, Teyssier12}]
by using  a similar technique that  the central density profiles of larger dwarf galaxies (i.e. M$\sim 10^7-10^8$ M$_\odot$) can  be flattened.  Reduced central
densities, however,  are not a general product  of simulations that include
gas physics.  Most galaxy profiles are either 
unchanged by feedback \cite{Parry12}  or made even steeper than those in
the dark-matter-only simulations due to  adiabatic contraction \cite{diCintio11}. The use of  hydrodynamical
sub-grid models may be responsible for
these divergent outcomes.  

Another approach invokes the partial removal of the baryon component from  gas-rich dwarf irregular galaxies by ram pressure.  This accounts for the observation that
the baryon fractions in dwarf spheroidal galaxies  are lower than the cosmic mean. When combined with the concentration of baryons in the inner part of the Milky Way halo, Arraki et al. \cite{Arraki14} could explain the circular velocities observed in Milky Way satellites [e.g. \cite{Penarrubia08, Strigari08, Lokas09, Walker09, Wolf10, Strigari10}] and resolve the discrepancy between observations and the $\Lambda$CDM model \cite{Zolotov12, Wang12}. 

\subsection{Resolution Problem}

We also  note that  dwarf
spheroidals generally present a  more difficult problem numerically in cosmological simulations 
because of  their small mass and physical size $\sim300$ pc. Hence,
it is difficult to resolve the central regions of dSph-size sub-halos  even for collision-less  simulations
\cite{Boylan-Kolchin12a}.  The mass  tends to be  systematically underestimated by 20\% within $4 - 5$ force resolution
elements because
of the gravitational softening inherent in the  simulations \cite{Font11}. 
Poor resolution can also give rise to two-body relaxation
errors that tend to flatten the inner density profile.
This  effect can propagate radially outward in the
cumulative velocity profile. If the dark matter potential
is artificially shallow due to poor  resolution,
then the gas outflow and tidal effect may over-predict the removal
of mass. This effect may be exacerbated by the time resolution of the simulations.  A time step that is too large can  lead to bursts of star formation, while dwarf galaxies may be better modeled as having slow and continuous star formation  \cite{Gallagher84, Krumholz14}.

These issues have motivated the use \cite{Garrison-Kimmel13} of controlled, idealized
simulations to achieve the required spatial, temporal,  and
mass resolution (1000 M$_\odot$). In that work it was  concluded that supernovae feedback is not  capable of solving the TBTF problem,
so  that problem remains.

\subsection{Puzzle of giant pure disk galaxies and the bulge of the Milky Way}
Our galaxy is an example of a pure-disk galaxy, one with no classical bulge of the sort that we associate with major mergers. M31 has a classical bulge, but even it has a bulge-to-total ratio of only about 1/4 in baryonic mass. A persistent puzzle \cite{Kormendy10, Kormendy12} has been how does hierarchical clustering form halos that have outer halos of nearly circular velocity without making a classical bulge?

 \section{Conclusion}
 In conclusion it is clear that there are a great many problems to be resolved before a detailed understanding of the origin and evolution of structure and nucleosynthesis in the Local Group can be obtained.  Although it is indeed encouraging that a consensus has emerged as to the dark matter dynamics and non-radiative gas dynamics, there is still much to be understood on the details of star formation and feedback by stellar winds, UV flux, Supernovae, AGNs, etc., as well as effects of dust, magnetic fields, radiation balance, and the nucleosynthesis and mixing of new elements from stars with the ISM, along with the history of galactic outflows and inflowing material.   No doubt the attempts to understand these processes will absorb many hours CPU and telescope time in the coming years.

\appendix 

\section{Links to various projects and codes}

For anyone interested in attacking the problems listed herein.  Here is a summary of some useful links to 
projects and resource codes.

\begin{table}[h]
\tbl{Links to various projects and codes}
{\begin{tabular}{cc}
 \toprule
   Project or code   & Link       \\
\colrule
AGORA Project &
https://sites.google.com/site/santacruzcomparisonproject/home \\

ERIS Project &
https://webhome.phys.ethz.ch/~jguedes/eris.html \\

Aquarius Project &
http://www.mpa-garching.mpg.de/aquarius/ \\

Aquila Project &
http://www.aip.de/People/cscannapieco/aquila/ \\

Millennium Simulation Project & http://www.mpa-garching.mpg.de/galform/virgo/millennium/ \\

MUSIC : Zoom Initial Conditions &
http://www.phys.ethz.ch/~hahn/MUSIC/ \\

AMIGA : Halo Finder &
http://popia.ft.uam.es/AMIGA/ \\

AGORA code  &
http://www.agorasimulations.org/  \\

GADGET-2 Code & http://www.mpa-garching.mpg.de/gadget/ \\

ENZO Code & code.google.com/p/enzo/? \\

Source Code Libraries &
http://www.astrosim.net/code/doku.php?id=home:home \\

-   &  ${\rm asterisk.apod.com/wp/?page_id=12}$   \\

\botrule
\end{tabular}}
\label{linktab} 
\end{table}

\section*{Acknowledgments}

 Work at the University of Notre Dame supported in part
by the U.S. Department of Energy under 
Nuclear Theory Grant DE-FG02-95-ER40934 and by the
Joint Institute of Nuclear Astrophysics (JINA) through NSF-PFC grant PHY08-22648.  This research was also supported in part by the University of Notre Dame Center for Research Computing.  Work in Vietnam supported in part by the Ministry of Education (MOE)
grant no. B2014-17-45.


\end{document}